\documentclass[aps,prl,superscriptaddress,twocolumn,nobibnotes]{revtex4-1}

\usepackage{graphicx}
\usepackage{todonotes}
\usepackage{amsmath}
\usepackage{amsmath}
\usepackage{upgreek}
\usepackage{color}

\usepackage{hyperref}

\renewcommand{\phi}{ \varphi }

\begin{document}

\title{Quantized Self-Assembly of Discotic Rings
in a Liquid Crystal Confined in Nanopores}

\author{Kathrin Sentker}
\affiliation{Institut f\"ur Materialphysik und -technologie, 
    Technische Universit\"at Hamburg, Ei{\ss}endorferstr. 42, D-21073 Hamburg, Germany}


\author{Arne W. Zantop}
\affiliation{Max-Planck-Institut f\"ur Dynamik und Selbstorganisation, 
    Am Fa{\ss}berg 17, D-37077 G\"ottingen, Germany}
\author{Milena Lippmann}
\affiliation{Deutsches Elektronen Synchrotron, 
	Notkestra{\ss}e 85, D-22607 Hamburg, Germany}

\author{Tommy Hofmann}
\affiliation{Helmholtz-Zentrum Berlin f\"ur Materialien und Energie,
Hahn-Meitner-Platz 1, D-14109 Berlin, Germany}

\author{Oliver H. Seeck}
\affiliation{Deutsches Elektronen Synchrotron, 
	Notkestra{\ss}e 85, D-22607 Hamburg, Germany}

\author{Andriy V. Kityk}
\affiliation{Faculty of Electrical Engineering, Czestochowa University of Technology, Al.~Armii Krajowej 17, P-42-200 Czestochowa, Poland}

\author{Arda Yildirim}
\affiliation{Bundesanstalt f\"ur Materialforschung und -pr\"ufung, Unter den Eichen 87, D-12205 Berlin, Germany}

\author{Andreas Sch\"onhals}
\affiliation{Bundesanstalt f\"ur Materialforschung und -pr\"ufung, Unter den Eichen 87, D-12205 Berlin, Germany}

\author{Marco G. Mazza}
\affiliation{Max-Planck-Institut f\"ur Dynamik und Selbstorganisation, 
    Am Fa{\ss}berg 17, D-37077 G\"ottingen, Germany}

\author{Patrick Huber}
\email[]{patrick.huber@tuhh.de}
\affiliation{Institut f\"ur Materialphysik und -technologie, 
    Technische Universit\"at Hamburg, Ei{\ss}endorferstr. 42, D-21073 Hamburg, Germany}

\date{\today}

\begin{abstract}
Disklike molecules with aromatic cores spontaneously stack up in linear columns with high, one-dimensional charge carrier mobilities along the columnar axes making them prominent model systems for functional, self-organized matter. We show by high-resolution optical birefringence and synchrotron-based X-ray diffraction that confining a thermotropic discotic liquid crystal in cylindrical nanopores induces a quantized formation of annular layers consisting of concentric circular bent columns, unknown in the bulk state. Starting from the walls this ring self-assembly propagates layer by layer towards the pore center in the supercooled domain of the bulk isotropic-columnar transition and thus allows one to switch on and off reversibly single, nanosized rings through small temperature variations. By establishing a Gibbs free energy phase diagram we trace the phase transition quantization to the discreteness of the layers' excess bend deformation energies in comparison to the thermal energy, even for this near room-temperature system. Monte Carlo simulations yielding spatially resolved nematic order parameters, density maps and bond-forientational order parameters corroborate the universality and robustness of the confinement-induced columnar ring formation as well as its quantized nature.
 \end{abstract}

\maketitle

Disklike molecules with aromatic cores and aliphatic side chains stack up in columns, which arrange in a two-dimensional lattice leading to discotic columnar liquid crystals (DLCs). Because of overlapping $\pi$~electrons of the aromatic cores DLCs exhibit long-range self-assembly and self-healing mechanisms in combination with high one-dimensional charge mobility along the columnar axes \cite{Adam1993, Oswald2005, Feng2009, Troisi2009, Sergeyev2007,Bisoyi2010,Kumar2010, Kumar2014,Woehrle2016}.

\begin{figure}[h!]
	\centering
	\includegraphics[width=0.45\textwidth]{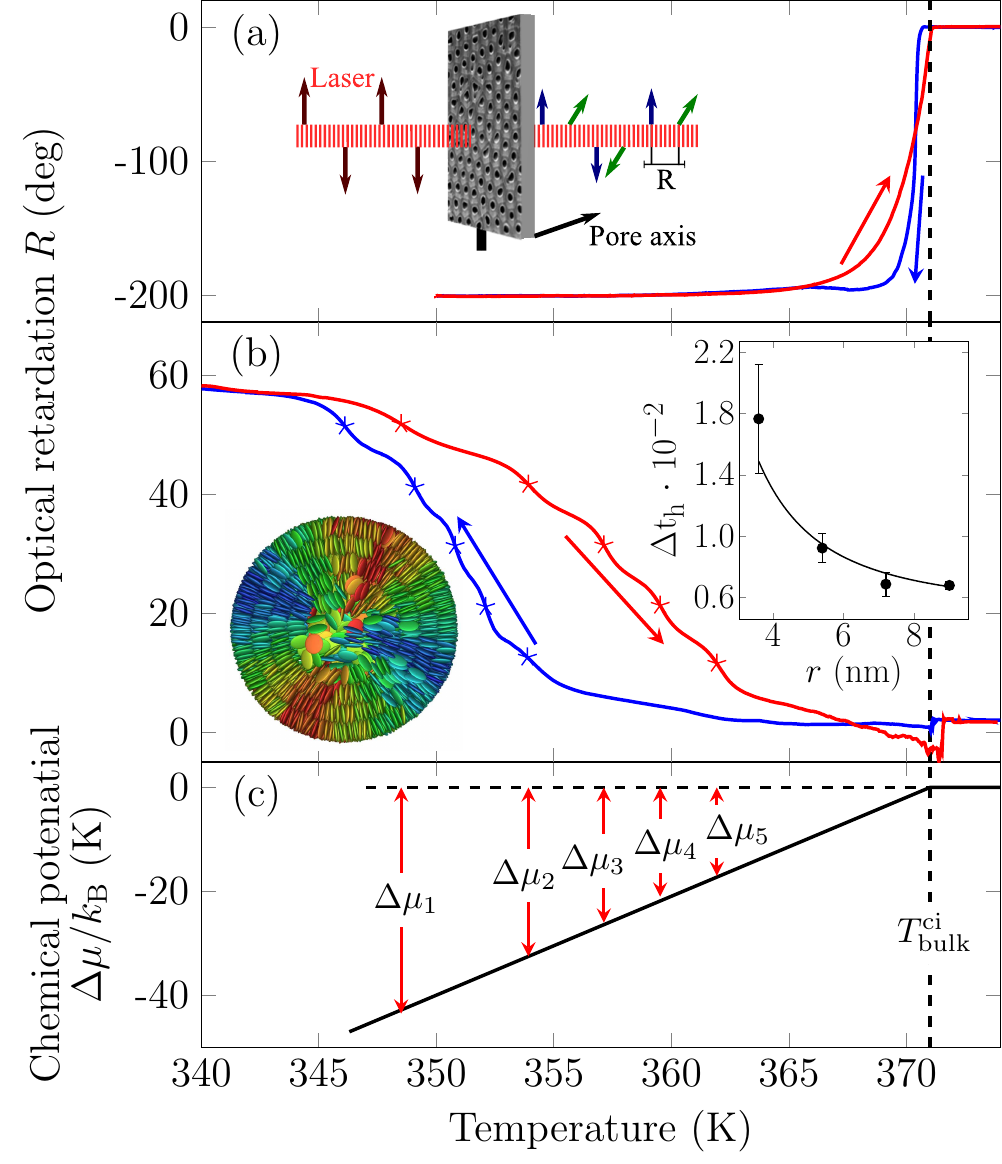}
	\caption{(Color online). (a) Birefringence experiment illustration and measured temperature evolution of the normalized retardation $R$ of HAT6 in the bulk and (b) confined state during cooling (blue) and heating (red). Insets: Normalized supercooling temperature for the isotropic-columnar transition of each annular layer with curvature radius $r$ along with a $r^{-2}$ fit. A Monte-Carlo simulation snapshot illustrating the formation of concentric columnar rings. The colors represent the relative orientation of the molecules with respect to the horizontal axis, where blue means 0$^{\circ}$ and red 90$^{\circ}$ alignment. (c) Chemical potential $\Delta \mu$- temperature $T$ phase diagram of HAT6 in the vicinity of the isotropic-columnar bulk transition. The excess energy $\Delta \mu_n$ of a molecule in ring $n$ is indicated by arrows at the supercooling $T_n$ ($n = 1 \dots 5$), marked by red asterisks in panel (b).}
	\label{fig:Measurement}
\end{figure}

These exceptional properties are strongly sensitive to interfacial interactions \cite{Charra1998, Oswald2005, Brunet2011, Duran2012, Calus2015,Huber2015, Ryu2016} having caused a broad interest in the behavior of DLCs in confined geometries, in particular with regard to their functionalities in organic electronics \cite{Schmidt-Mende2001,Steinhart2005,Brunet2011,Duran2012,Kityk2014, Kumar2014,Ryu2016, Zhang2017}. Recently, it was reported that nanopore-confined DLCs can form concentric supermolecular ring structures, absent in the bulk \cite{Zhang2015}. However, there is still little knowledge about this self-assembly due to challenges in resolving orientational order at interfaces \cite{Lee2009,Xia2015}, a lack of temperature-dependent formation studies and the complex interplay of interfacial interactions and pure confinement effects in nanoscopic systems \cite{Christenson2001, Alba-Simionesco2006, Binder2008, Huber2015}.

Here we present a temperature-dependent optical birefringence, X-ray diffraction and Monte Carlo simulation study on the structure of an archetypical DLC (HAT6) confined in an array of cylindrical pores ($17$~nm across, $360~\upmu$m in length) in a silica membrane. It is aimed at understanding the thermodynamics and the structural evolution of the isotropic-columnar transition in this extreme spatial confinement.

A suitable technique to study orientational order of DLCs is an optical birefringence measurement, see Fig.~\ref{fig:Measurement}(a) and Ref.~\cite{Kityk2014} as well as the Supplemental Material \cite{supplemental} with Refs.~\cite{Sailor2011, Canham2015,Gruener2008a, Gruener2011} for the sample preparation. The optical retardation $R$ between the perpendicularly polarized ordinary and extraordinary beams is a direct measure of the orientational order within the sampling volume \cite{Kityk2008, Calus2014, Kityk2009}. 

As a reference in Fig.~\ref{fig:Measurement}(a) the $T$-dependence of the retardation $R(T)$ of bulk HAT6 embedded in a glass cell with a $10~\upmu$m gap is shown. Starting from the isotropic phase the sample undergoes a cooling-heating cycle at a rate of $0.03$~K/min. Upon cooling, $R(T)$ exhibits a drastic drop from 0, typical of the disordered isotropic liquid, to negative values at the isotropic-columnar transition $T^{\text{ci}}_{\text{bulk}} = 371$~K indicating the formation of a face-on orientation at the glass surface and thus column formation along the surface normal. Upon heating, $R(T)$ vanishes at $T^{\text{ci}}_{\text{bulk}}$. 

Figure~\ref{fig:Measurement}(b) shows the $R(T )$ of HAT6 shows R(T) of HAT6 imbibed in nanopores with edge-on molecular anchoring at the walls, which is achieved by a silanization of the silica walls and thus via replacement of hydrophilic hydroxyl groups with hydrophobic methyl groups. In contrast to the bulk case, $R(T)$ increases towards positive values upon cooling (0.15~K/min) indicating alignment of the molecular director $\boldsymbol{\hat{n}}$ perpendicular to the pore axis, see the inset in Fig.~\ref{fig:Measurement}(b). Interestingly, the collective molecular order does not evolve in a monotonic manner. Rather, a sequence of small plateaus, separated by five pronounced changes in $R(T)$, results in a staircaselike transition, both upon cooling and heating. This behavior is reproducible in cooling-heating cycles.

The distance of the \{10\} hexagonal planes in the columnar bulk phase is $d_{\text{cc}}=1.8$~nm \cite{Krause2014, Zhang2015}, fitting roughly 10 times into the pore diameter. Cooling down from the isotropic phase the molecules closest to the pore wall start to orient with their director parallel to the pore wall, but still perpendicular to the pore axis, forming, due to the cylindrical confinement, a bent columnar concentric ring, see Fig.~\ref{fig:Measurement}(b). Analogous observations have been made for DLC confined in larger nanopores \cite{Zhang2014,Zhang2015, Zhang2017}. With decreasing $T$ this order propagates to the center forming five concentric columnar rings with increasing ring curvature, see inset in Fig.~\ref{fig:Measurement}(b). The formation of each layer contributes to the increase of $R(T)$ at distinct $T$s as marked in Fig.~\ref{fig:Measurement}(b). 

A pronounced cooling-heating hysteresis is present. In contrast to cooling, where random nucleation processes delay column formation, upon heating the disordered high-$T$ phase is nucleated in the center \cite{Kityk2014}. There the largest curvature and geometric frustration in the low-$T$ phase favoring the isotropic phase occur. The isotropic phase expands layer by layer toward the wall, resulting in a quantized decrease in $R(T)$. Because of the laser beam's final size, $R(T)$ corresponds to an averaging of the molecular orientation over multiple nanopores, where geometric randomness can lead to variations in the layer transition $T$s. Thus, $R(T)$ does not appear with sharp but rather smeared transition points. 

It is surprising, that the columns do not align axially to fulfill the edge-on anchoring even without the necessity of bent columns. However, as outlined in Ref.~\cite{Zhang2015} this results in substantial excess elastic energies originating from the distortions of the 2D hexagonal column lattice at the curved pore surfaces.

\begin{figure}
	\centering
	\includegraphics[width=0.45\textwidth]{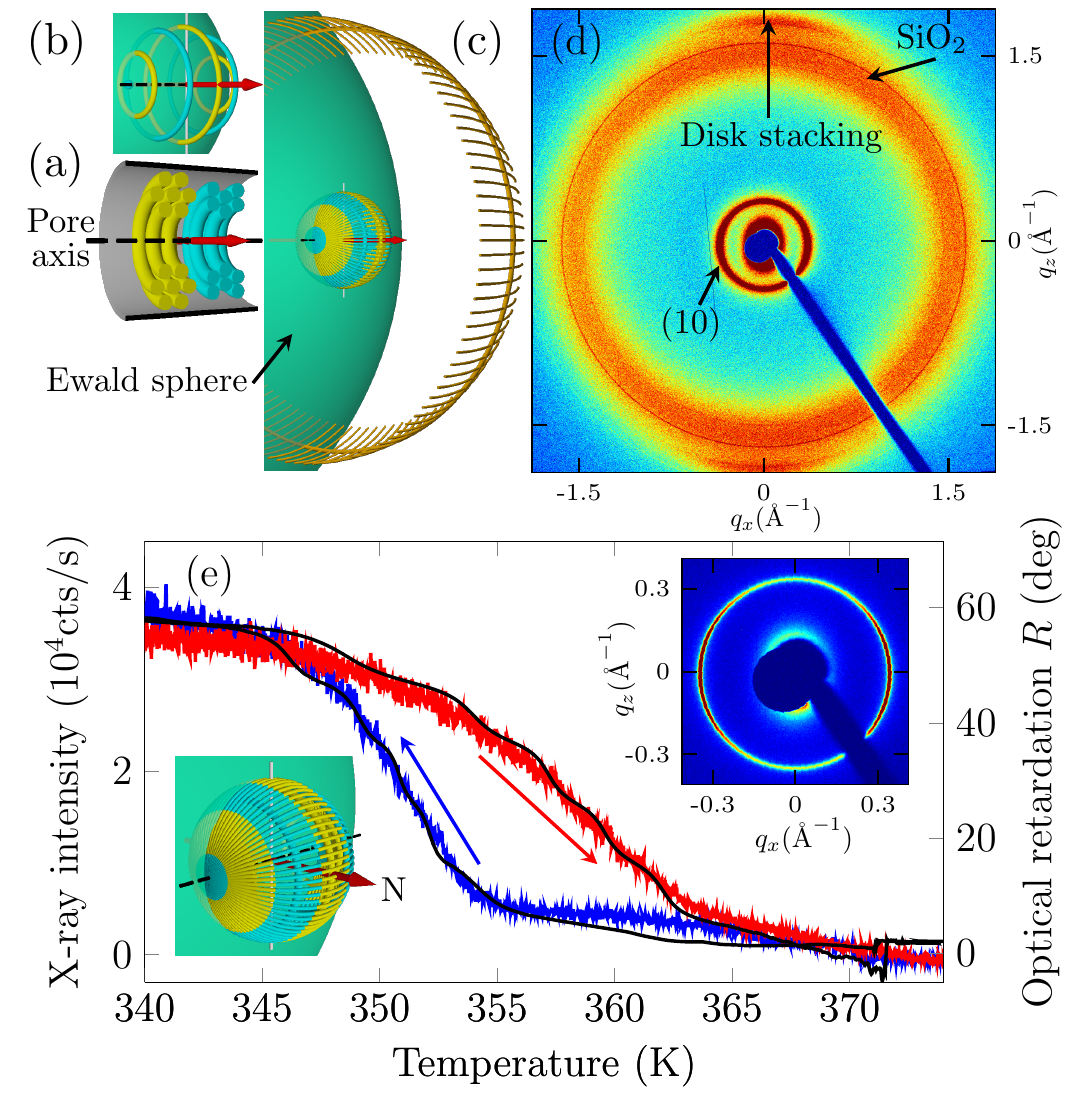}

	\caption{(Color online) (a) Circular columnar domains with \{10\} (yellow) and \{11\} (green) wall orientation in a cylindrical pore and reciprocal space maps ($\omega =$90$^{\circ}$) assuming (b) perfect aligned and (c) a randomization of the domain orientations by maximal 15$^{\circ}$ with regard to the mean pore axis direction. (d) X-ray diffraction pattern of HAT6 (T=340~K, $\omega =$ 75$^{\circ}$) confined in nanopores. (e) Temperature evolution of the (10) Bragg ring for a cooling-heating cycle. $R(T)$ from Fig.~\ref{fig:Measurement}(b) serves as a guide to the eye. Insets: Enlarged reciprocal space and diffraction pattern focusing on the (10) Bragg ring.}
	\label{fig:XRD}
\end{figure}

A $T$-dependent X-ray diffraction experiment, sensitive to translational order in cross sections aligned parallel to the long pore axes ($\omega$= 75$^{\circ}$) is performed at the P08 beamline \cite{Seeck2012} of the PETRA III synchrotron, see Fig.~\ref{fig:XRD}(a) and supplemental. Upon cooling and heating we observe the appearance and vanishing of an intensity ring at a wave vector transfer $q_{\rm (10)}=$ (0.3445 $\pm$ 0.0001)~\AA$^{-1}$, typical of the (10) Bragg reflection of hexagonal intercolumnar order, as well as two streaks in the equatorial directions at $q_{\rm dd}= $(1.7726 $\pm$ 0.008)~\AA$^{-1}$, characteristic of the intracolumnar disk-disk stacking, see Fig.~\ref{fig:XRD}(d). Note that we follow here standard texture analysis nomenclature, so that the equator is in vertical, whereas the poles are in the incident beam (horizontal) direction. As illustrated in the ideal reciprocal space map (Fig.~\ref{fig:XRD}(c), $\omega$= 90$^{\circ}$), the two equatorial intensity streaks represent the quasi-Bragg peaks resulting from a cutting of the Ewald sphere into the radially aligned Bragg ring of intracolumnar stacking. 

To explain the (10) ring we anticipate a coexistence of \{10\} and \{11\} domain orientations, leading to a 12-fold diffraction pattern, when the Ewald sphere cuts in the corresponding (10) rings, see Fig.~\ref{fig:XRD}(b) and the Supplemental Material for such a texture measured for HAT6 in larger channels. Additionally, a randomization of the domain orientations (and columnar ring orientation) with regard to the averaged pore direction of at least 15$^{\circ}$, in agreement with the large azimuthal width of the intrastacking peaks at large $q$, leads to an apparent isotropization of the (10) orientations with a densification towards the pore axis (polar) directions, see the inset in Fig.~\ref{fig:XRD}(e). As the Ewald sphere cuts into the resulting (10) sphere, a Bragg ring with azimuthal intensity maxima in the polar (horizontal) directions is expected, in agreement with our observation. Likely the randomness of the domain and ring orientation originates from a sizable tortuosity (meandering) of single nanopores, similarly as inferred from capillary filling experiments on the untransformed silicon nanopores \cite{Acquaroli2011}. Note that a ''logpile'' texture, where straight columns cross the pores perpendicularly with random orientation, as has been found for cylindrical pores and planar wall anchoring \cite{Zhang2014}, would also result in a (10) diffraction ring, albeit it would be incompatible with the optical experiments, the simulation study and the edge-on pore wall anchoring.

The $T$-dependent (10) Bragg ring intensity, see Figure~\ref{fig:XRD}(e), follows remarkably well the birefringence, both in the onset and in the hysteresis width. However, the staircase behavior is less pronounced. The Monte Carlo simulations discussed below suggest that this originates in defect healing or formation in the hexagonal order of the already or still present rings which continuously occur upon cooling and heating, respectively, and to which the birefringence is insensitive. In contrast, they lead to continuous (10) intensity changes, additionally to the stepwise changes upon layer formation or vanishing, and thus to a stronger temperature smearing of the diffraction compared to the optical signal.

To analyze the thermodynamics of the confined system with respect to the bulk one we plot a chemical potential $\Delta \mu$- temperature $T$ phase diagram \cite{Huber1999} close to the bulk transition, see Figure~\ref{fig:Measurement}(c). As a reference the bulk isotropic liquid $\mu^{\rm iso}_{\rm bulk}$ (solid line) and its metastable extension $T < T^{\rm ci}_{\rm bulk}$ (dashed line) are plotted at $\Delta \mu =$ 0. The chemical potential of the bulk columnar phase intersects with $\mu^{\rm iso}_{\rm bulk}$ at $T^{\rm ci}_{\rm bulk}$. Its slope is given by the entropy change $\Delta S$ between the isotropic and the columnar phase. From measurements of the latent heat $H$ of the isotropic-columnar transition we determine $\partial \mu^{\rm col}_{\rm bulk}/\partial T = - \Delta  S = H/T^{\rm ci}_{\rm bulk}$. Chemical potentials per molecule in the $n$th ring calculated by $\Delta \mu^{\rm col}_{\rm n} = \Delta t_{\text{th},n} \cdot H \cdot m_{\text{mol}}/(N_A k_B)$ with $m_{\text{mol}}$ the molar mass of HAT6, $N_A$ the Avogadro constant, $k_{B}$ the Boltzmann constant, $\Delta t_{\text{th},n} = (T_{n} - T^{\rm ci}_{\rm bulk})/T^{\rm ci}_{\rm bulk}$, mark the energy differences between the metastable liquid and the columnar phase. The transition in each annular layer $n$ occurs (red asterisks in Fig.~\ref{fig:Measurement}~(b)) when the excess energy due to the confinement is balanced by the corresponding supercooling energies. 

Neglecting lattice distortions at the pore and domain walls, the dominant mechanism contributing to the excess energy of the rings (and thus to the supercooling) is given by the strong bend of the columns. Thus, the Frank bend elastic energy density per unit length $f_\text{B}$ with constant $K_{3}$ should contribute significantly. For ring $n$ with radius $r_n$ it reads $f_\text{B} = \frac{K_{3}}{2} (\boldsymbol{\hat{n}}\times\nabla\times\boldsymbol{\hat{n}})^2 = \frac{K_{3}}{2}  r_{n}^{-2}$ and the corresponding supercooling $T$ differences between subsequent rings are given by $\Delta t_{\text{th},i}= (T_{i}-T_{i+1})/T^{\rm ci}_{\rm bulk} = K_{3}/(2   H   \rho_{\text{HAT6}})   r_{i}^{-2}$ with $i =  1 \dots 4$. These $T$s plotted versus ring radii $r$ are shown in the inset of Fig.~\ref{fig:Measurement}(b) along with a fit according to the equation derived above. A good agreement with an $r^{-2}$ scaling yielding the bend elastic constant of $K_3$ = (2.7 $\pm$ 0.7)~pN is found, a value in reasonable agreement with the one reported for the chemically closely related HAT7 $K_3 = 4$~pN \cite{Sallen1995}.

To obtain a microscopic picture, we perform parallel-tempering Monte Carlo simulations of $N$ DLC molecules in the isothermal-isobaric ensemble expanded by temperature, see the Supplemental Material, which includes 
Refs.~\cite{caprion2009discotic,swendsenPRL1986,Yan1999,Earl2005,Lechner2008}. We employ the Gay-Berne-II model for the DLC\cite{BatesJCP1996,Caprion2003}. To analyze the structure of the model fluid we define an average order parameter $\bar{S}=\langle {\frac{1}{N} \sum_i S_\mathrm{loc} |_{\mathsf{B}(i)}} \rangle$ for $i=1,\dots, N$, where we calculate the local nematic order in a sphere $\mathsf{B}(i)$ with radius $2.5\; \sigma_\text{ff}$ centered around particle $i$. The angle brackets $\langle\dots\rangle$ indicate an ensemble average. 

\begin{figure}
	\centering
         \includegraphics[width=0.45\textwidth]{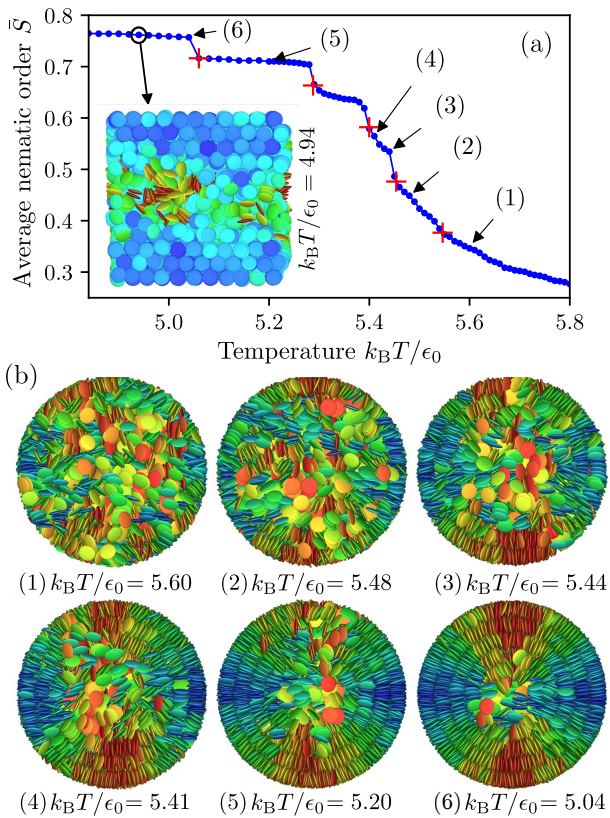}
	\caption{(Color online). (a) The temperature-dependence of the average nematic order parameter $\bar{S}$ (blue circles) shows discontinuous jumps, marked with red crosses. Inset: Cross-sectional view on the molecular arrangement in the nanopore at a temperature $k_\mathrm{B}T/\epsilon_0=4.94$. (b) Radial snapshots of molecular configurations at the $T$s marked as ($1\dots 6$) in panel (a).} 
	\label{fig:local_nematic}
\end{figure}

Figure~\ref{fig:local_nematic}(a) shows the $T$-dependence of $\bar{S}$. Similar to the optical experiments, a stepwise increase in the orientational order is clearly visible as $T$ decreases. Typical molecular configurations at different $T$s are shown in Fig.~\ref{fig:local_nematic}(b).

\begin{figure}
	\centering
	\includegraphics[width=0.9\columnwidth]{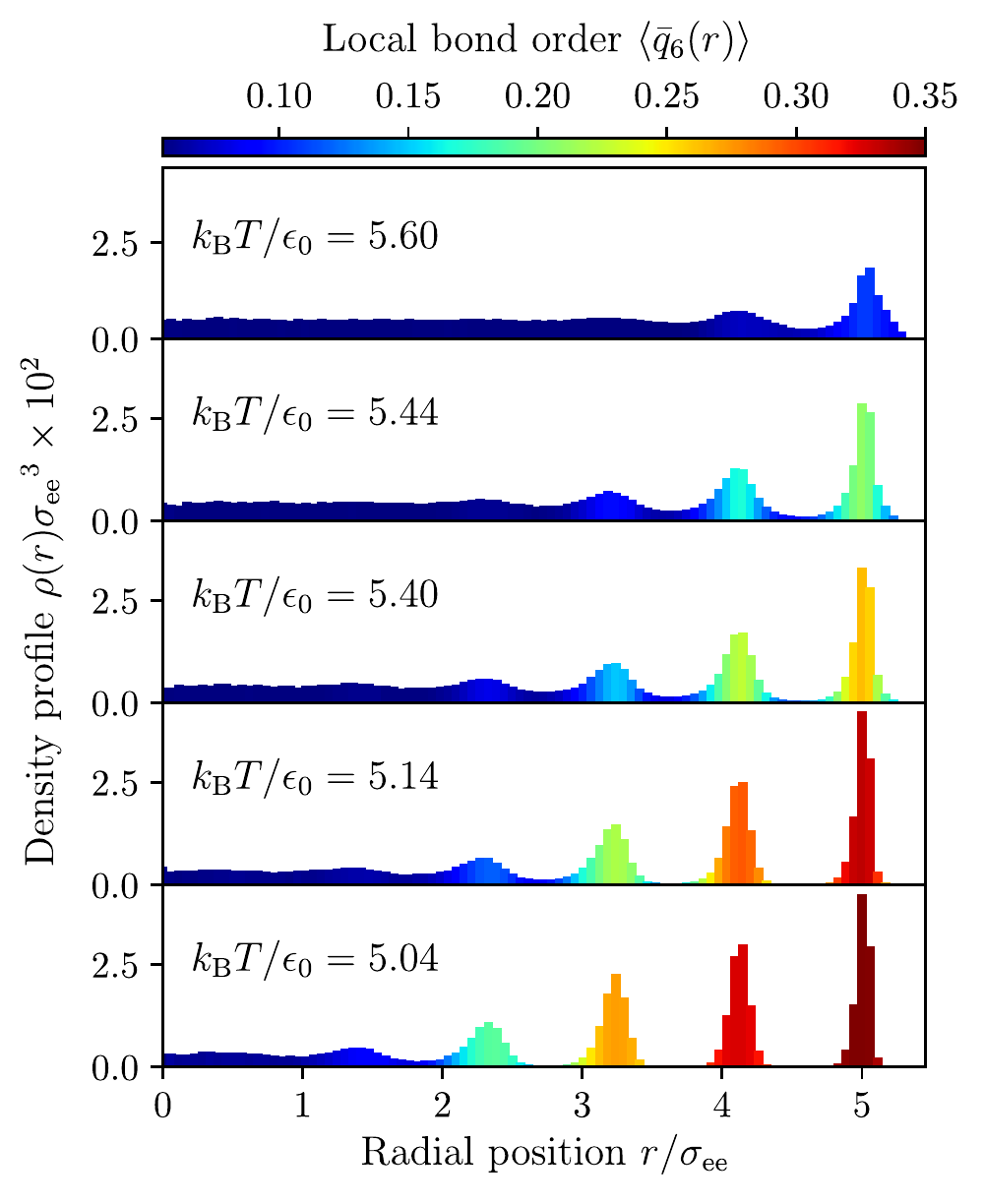}
	\caption{(Color online). Radial dependence of the local density $\rho(r)$ and bond orientational order parameter $\bar{q}_6$ at different temperatures along the quantized phase transition.}
	\label{fig:bond_orientational_order}
\end{figure}

Upon cooling, the fluid undergoes two symmetry breakings: broken translational invariance that manifests itself with a periodic density modulation of the molecules' centers of mass, and a broken rotational symmetry that singles out specific molecular orientations. Because of the liquid-crystalline nature of the fluid, these two symmetries are intimately connected: as a concentric ring emerges, the orientations within it are strongly correlated. We quantify both symmetry breakings by calculating the local bond order parameter $\bar{q}_6$, and the radial density profile $\rho(r)\equiv \langle \frac{1}{\pi r^2h\rho_0} \frac{1}{N} \sum_i\delta(r_i-r)\rangle$, where $r_i$ is the distance of the molecule from the pore axis, $\rho_0$ the average density and $h$ the height of the confining cylinder.

Figure~\ref{fig:bond_orientational_order} shows both $\rho(r)$ and $\bar{q}_6(r)$ at five different $T$s.  At high $T$ ($k_\mathrm{B}T/\epsilon_0=5.6$), a paranematic state is present, see Fig.~\ref{fig:local_nematic}(b). Upon cooling below $k_\mathrm{B}T/\epsilon_0=5.52$, a first circular configuration emerges. As the $T$ decreases below $k_\mathrm{B}T/\epsilon_0=5.48$, a second ring forms, see the large value of $\bar{q}_6$. Below the third transition at $k_\mathrm{B}T/\epsilon_0=5.42$, a third circular concentric configuration appears. We note that the bond orientational order of the rings already present continues to grow, with widespread disappearances of defects, supporting our interpretation of the differences in the X-ray and optical experiment. 
The orientational order consistently grows upon cooling below the fourth transition at $k_\mathrm{B}T/\epsilon_0=5.31$.
As $T$ decreases below the fifth transition at $k_\mathrm{B}T/\epsilon_0=5.06$, strong orientational order permeates the system, except in the pore center, which remains a defect region. Thus, upon cooling, particles become much more localized within circular concentric configurations and form hexagonal columnar arrangements, see inset in Fig.~\ref{fig:local_nematic}(a). After each formation of a new ring, the density increases in the previous layers. At the lowest $T$s, five regions of enhanced density are clearly visible, corresponding to the circular concentric configurations. 

\begin{figure}
	\includegraphics[width=0.9\columnwidth]{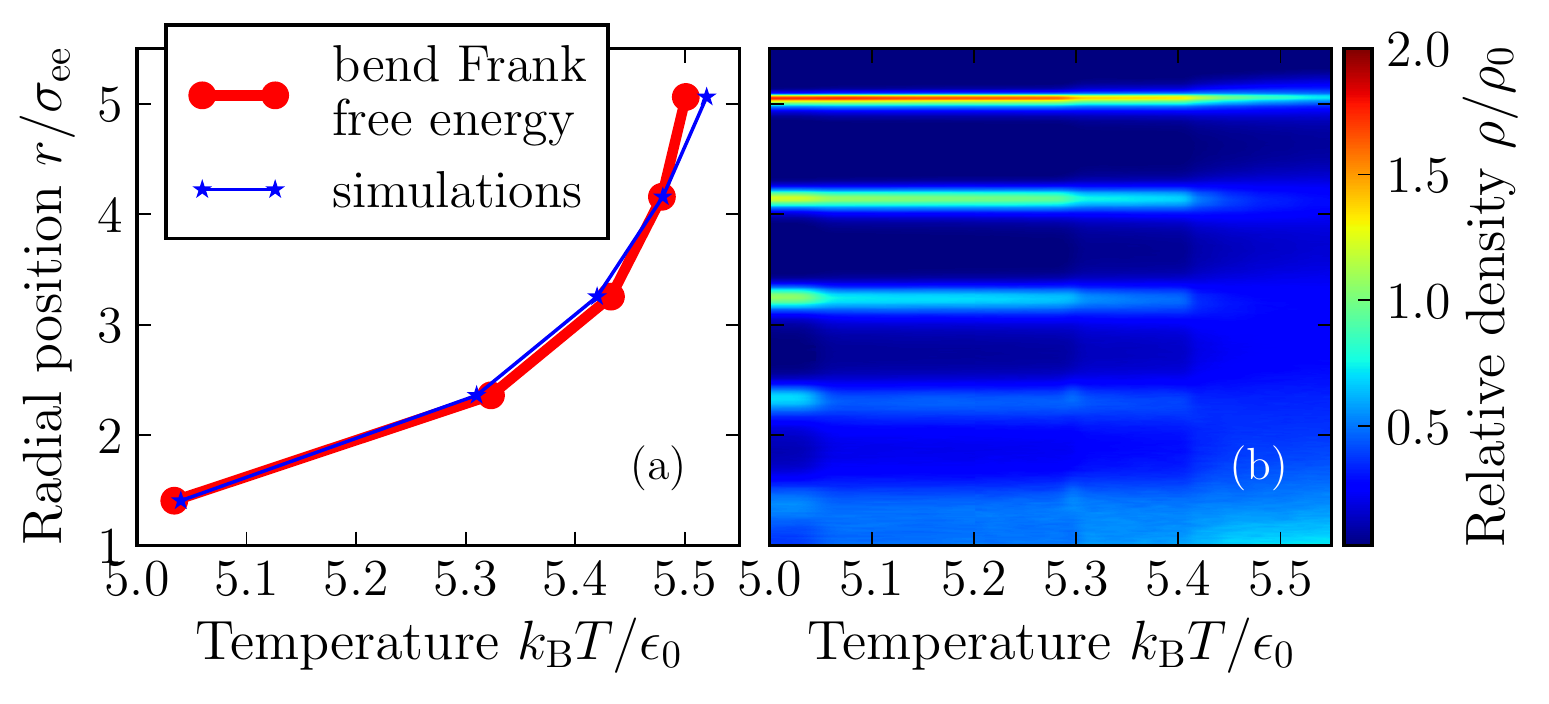}
\caption{(Color online). (a) Dependence of the transition temperatures on the ring's radial position.  (b) The temperature evolution of the density profile. Upon cooling, five distinct regions of large density appear, corresponding to the circular concentric layers.}
\label{fig:density}
\end{figure}

The simulations also support the picture that the ring curvatures cause the quantized transition. Figure~\ref{fig:density}(a) shows the dependence of the transition $T$s on the radial ring distance, as obtained from the $T$-evolution of the density profile, see Fig.~\ref{fig:density}(b). The pore center constitutes a defect with a strong gradient of local nematic order. Therefore a term proportional to $(\nabla S)^2$ and the dependence $K\propto S^2$ are included in the calculation of the Frank free energy. We find that the bend Frank free energy prediction of an $r^{-2}$-scaling is in good agreement with the simulation.

As seen in the simulation, well-defined hexagonal order is a prerequisite for discrete layer curvatures and the resulting layer-by-layer growth in the $<$11$>$ direction. Our experiments on an array of nanopores are, however, affected by local and global geometry variations. Single-pore experiments could infer in the future, how these different disorder contributions result in a smearing of the observed quantization compared to the theoretical expectation.
 
In summary, we have found a quantized phase transition of a liquid crystal confined to nanopores. Optical birefringence, X-ray diffraction experiments, together with Monte Carlo simulations show that the stepwise transformation originates in the formation of circular concentric rings. This finding is reminiscent of the quantized nature of the isotropic-smectic transition, previously reported for rodlike mesogens at planar interfaces \cite{Ocko1986,Selinger1988, Ocko1990} and discrete layer formation in physisorbed or colloidal systems at planar surfaces \cite{Bruch2007, Born2014}. Whereas there the energy scale for the quantization is set by interfacial interactions, here it is determined by the discreteness of the layers' excess bend deformation energies in comparison to the thermal energy.

The phase transition quantization exemplifies in a remarkable manner how confinement can alter the physics of liquid crystals \cite{Crawford1996, Berreman1972,Sheng1976, Stark2004, Kityk2008, Jin2003,Fukuto2008, Binder2008, Fukuda2011, Kityk2014, Calus2014, Huber2015, Jeong2015, Schlotthauer2015, Dietrich2017, Kim2017, Busch2017} and allows us to determine the otherwise hard to access bend elastic constant \cite{OndrisCrawford1993}. More generally, it highlights how curved geometries can alter self-assembly and crystallization  \cite{Christenson2001, Bausch2003, Alba-Simionesco2006, Muter2010, Huber2015} and how versatile soft matter can adapt to extreme spatial constraints with new architectural principles dynamics, as has been similarly discussed for simple molecular \cite{Huber1999, Huber2004, Franosch2012, Schappert2013} and polymeric systems \cite{Yu2006, Martin2010,Kusmin2010, Carvalho2015, Egorov2016, Nikoubashman2017}. Finally, we envision that the spontaneous, temperature-tunable nanoscale ring formation demonstrated here along with the one-dimensional charge carrier pathways and mechanical stability of the membranes may provide a versatile playground for the study of electronic and magnetic confinement effects \cite{Lorke2000} or even of the fluid-wall-interaction-induced deformations of nanopores \cite{Gor2017}. It may also serve as nanotemplating mechanism for organic semiconductor-based devices \cite{Lorke2000,Steinhart2005, Duran2012, Woehrle2016} given the nowadays readily available nanoporous solids  \cite{Sailor2011, Furukawa2013, Kuster2014,Canham2015,Liu2017, Juarez2017} and the simple preparation by capillarity-driven, spontaneous melt imbibition\cite{Gruener2011}. 

\begin{acknowledgments}
A.W.Z. and M.G.M. gratefully acknowledge Max Planck Society for funding, and the Deutsche Forschungsgemeinschaft (SFB 937, project A20) for support. A.V.K. acknowledges funding from the European Union's Horizon 2020 research and innovation program under the Marie Sk\l{}odowska-Curie grant agreement No. 778156. The German Science foundation contributed by the project SCHO 470/21-1 and HU 850/5-1 "Discotic Liquid Crystals in Nanoporous Solids: From the Structure and Dynamics to Local Charge Transport". P.H. and K.S. profited from the support within the Collaborative Research Initiative SFB 986, ÒTailor-Made Multi-Scale Materials SystemsÓ, projects B7, Z3, Hamburg (Germany). We thank Deutsche Elektronen Synchrotron DESY, Hamburg for access to the beamline P08 of the PETRA III synchrotron. We thank Christian Bahr, Stephan Herminghaus, and Sergej P\"uschel-Schlotthauer for insightful discussions and comments. K.S. and A.W.Z. contributed equally to this work. 

\end{acknowledgments}

\end{document}